\documentclass[12pt,amsmath,amssymb,nofootinbib,superscriptaddress]{revtex4-1}
\usepackage{amsmath} %
\usepackage{pstricks}
\usepackage{color} 
\usepackage{amssymb}
\usepackage{slashed}
\usepackage{graphicx,amsfonts,layout,appendix,subfigure}
\input{epsf} 
\def\beq{\begin{equation}}
\def\eeq{\end{equation}}
\def\bea{\begin{eqnarray}}
\def\eea{\end{eqnarray}}
\def\beqn{\begin{eqnarray}} 
\def\eeqn{\end{eqnarray}}

\def\nn{\nonumber}
\def\Eq#1{Eq.~(\ref{#1})}
\def\ln#1{\mathrm{ln}\left(#1\right)}
\def\lnn#1{\mathrm{ln^2}\left(#1\right)}
\def\li#1{\mathrm{Li_2}\left(#1\right)}

\newcommand\alphas{\alpha_{\mathrm{S}}}
\newcommand\as{a_{\mathrm{S}}}

\def\beq{\begin{equation}} \def\eeq{\end{equation}}
\def\beqn{\begin{eqnarray}} \def\eeqn{\end{eqnarray}}
 \def\to{\rightarrow}
\def\nn{\nonumber}

\begin{document} 

\newcommand\sss{\scriptscriptstyle}
\newcommand\IDC{\textbf{\textit{{\rm Id}}}_C}
\newcommand\SUNT{\textit{\textbf{T}}}
\newcommand\pslashed{\slashed{p}}
\newcommand\DST{D_{\rm ST}}
\newcommand\DDirac{D_{\rm Dirac}}
\newcommand\factor{\rm Factor}
\newcommand\Spmatrix{\textit{\textbf{Sp}}}
\newcommand\Spelement{{\rm Sp}}
\newcommand\Spfunction{{\rm Split}}
\newcommand\NC{N_C}
\newcommand\CG{c_{\Gamma}}
\newcommand\DR{D_{\rm R}}
\newcommand\aem{\alpha_{\rm em}} \newcommand\refq[1]{$^{[#1]}$}
\newcommand\avr[1]{\left\langle #1 \right\rangle}
\newcommand\lambdamsb{\Lambda_5^{\rm \sss \overline{MS}}}
\newcommand\MSB{{\rm \overline{MS}}} \newcommand\MS{{\rm MS}}
\newcommand\DIG{{\rm DIS}_\gamma} \newcommand\CA{C_{\sss A}}
\newcommand\DA{D_{\sss A}} \newcommand\CF{C_{\sss F}}
\newcommand\TF{T_{\sss F}} 
\newcommand\qeps{q^2_{\epsilon}} 

\begin{titlepage}
\renewcommand{\thefootnote}{\fnsymbol{footnote}}
\begin{flushright}
     ICAS 03/15\\ IFIC/15-81
     \end{flushright}
\par \vspace{10mm}

\begin{center}
{\Large \bf
QED corrections to the Altarelli-Parisi splitting functions
}
\end{center}

\par \vspace{2mm}
\begin{center}
{\bf Daniel de Florian}~$^{(a,b)}$\footnote{{\tt deflo@df.uba.ar}},
{\bf Germ\'an F. R. Sborlini}~$^{(c)}$\footnote{{\tt gfsborlini@df.uba.ar}}
 and 
{\bf Germ\'an Rodrigo}~$^{(c)}$\footnote{{\tt german.rodrigo@csic.es}}

\vspace{5mm}

${}^{(a)}$Departamento de F\'\i sica and IFIBA, FCEyN, Universidad de Buenos Aires, \\
(1428) Pabell\'on 1 Ciudad Universitaria, 
Capital Federal, Argentina

\vspace*{2mm}
${}^{(b)}$
International Center for Advanced Studies (ICAS), UNSAM, Campus Miguelete, 25 de Mayo y Francia,
(1650) Buenos Aires, Argentina

\vspace*{2mm}
${}^{(c)}$Instituto de F\'{\i}sica Corpuscular, 
Universitat de Val\`encia - Consejo Superior de Investigaciones Cient\'{\i}ficas,
Parc Cient\'{\i}fic, E-46980 Paterna, Valencia, Spain \\

\vspace{5mm}

\end{center}

\par \vspace{2mm}
\begin{center} {\large \bf Abstract} \end{center}
\begin{quote}
\pretolerance 10000

We discuss the combined effect of QED and QCD corrections to the evolution of parton distributions.
We extend the available knowledge of the Altarelli-Parisi splitting functions to one order higher in QED, 
and provide explicit expressions for the splitting kernels up to ${\cal O}(\alpha \, \alphas)$. The results presented in this article allow to perform a parton distribution function analysis reaching full NLO QCD-QED combined precision.

\end{quote}
\vspace{1cm}
\begin{center}
This paper is dedicated to the memory of Guido Altarelli
\end{center}
\vspace*{\fill}
\begin{flushleft}

May 2016

\end{flushleft}
\end{titlepage}

\setcounter{footnote}{0}


\setcounter{footnote}{0}
\renewcommand{\thefootnote}{\fnsymbol{footnote}}

\section{Introduction}
\label{sec:introduction}
With the advent of the Run II of the Large Hadron Collider (LHC), a large number of processes will be probed within a formidable accuracy. The precision reached at the experimental level needs to be matched from the theoretical side, in order to both extract information on the Standard Model (SM) parameters and identify possible effects for Beyond the Standard Model (BSM) physics. 

Theoretical predictions at the LHC require the convolution of the so-called partonic cross-sections and the parton distribution functions (PDF), providing information on the hard and soft aspects of the process, respectively. 

From the point of view of the partonic cross-sections, Next-to-Leading-Order (NLO) corrections in QCD (${\cal O}(\alphas)$) are the baseline for any realistic study and, thanks to a number of tools developed during the last decade, today it is possible to reach this accuracy in an almost automatic way. The state-of-the-art in fixed order computations for processes with up to two hard partons in the final state is reaching Next-to-Next-to-Leading-Order (NNLO), i.e. ${\cal O}(\alphas^2)$. 
The theoretical work includes not only fixed order calculations but also the application of resummation techniques, the implementation of more elaborated parton showers, and their matching with fixed order calculations, among other developments.

Given that $\alphas^2\sim \alpha$, it becomes necessary to include also the corresponding NLO ElectroWeak (EW) corrections, that for many observables, and particularly in the case of the production of particles with very large transverse momentum, exceed the few percent level and become quantitatively very important for an accurate description. 

Despite of the great achievements from the perturbative side, the situation would not be fully satisfactory without the corresponding improvements on the non-perturbative sector. On one hand, the splitting functions that run the evolution of parton distributions are known at NNLO in QCD \cite{Moch:2001im,Moch:2004pa,Vogt:2004mw,Vogt:2005dw}. On the other, the development of new global parton distribution analysis that incorporate the state-of-the-art in the evaluation of cross-sections and a larger amount of precise experimental data for many different observables, allows for a better estimate of statistical uncertainties on top of the central sets \cite{Alekhin:2013nda,Dulat:2015mca,Jimenez-Delgado:2014twa,Abramowicz:2015mha,Harland-Lang:2014zoa,Ball:2014uwa}.
Recent work has been done on the PDF sector to incorporate the EW effect (strictly speaking the dominant QED terms) in the evolution equations \cite{Martin:2004dh,Ball:2013hta,Bertone:2013vaa}.  The first significant change in the evolution of parton distributions with QED corrections is the appearance of a new distribution: the photon density (and eventually the leptonic densities as well). So far, only LO QED kernels were known to perform the evolution of parton densities \footnote{One-loop triple collinear splitting functions with photons have been recently computed in \cite{Sborlini:2014mpa,Sborlini:2014kla}.}.

Modern analysis, performed up to NNLO in QCD and LO in QED show that the photon PDF contribution is not negligible and needs to be carefully studied for precise predictions at the LHC, and even more for higher energies as the FCC-hh \cite{Schmidt:2015zda,Sadykov:2014aua,Carrazza:2015dea,Bertone:2015lqa}. On the other hand, the contribution from lepton PDFs is usually extremely suppressed. As stated, those analysis only include QED contributions to the lowest order, since the NLO combined QCD-QED contributions (i.e., ${\cal O}(\alpha \, \alphas)$) were not available. 

The main purpose of this paper is to present, for the first time, explicit expressions for the Altarelli-Parisi splitting kernels \cite{Altarelli:1977zs} to ${\cal O}(\alpha \, \alphas)$. We compute them by profiting from the original calculation of the NLO QCD corrections performed in Refs. \cite{Curci:1980uw,Furmanski:1980cm,Ellis:1996nn}, conveniently modifying the colour structures for each topological contribution. 
We explicitly concentrate on the QED corrections, without including those arising from Weak bosons, which only become relevant for extreme kinematical conditions.

Concerning hadronic cross-sections, a full NNLO contribution in the context of QCD$+$QED requires the knowledge of the kernels presented in this paper to perform the subtraction of IR singularities and define the corresponding factorization scheme at this order. Furthermore, until the full 3-loop (mixed QCD-QED) splitting functions become available, they will be essential to evolve the parton distributions to a higher accuracy than the one available so far. 

The structure of the manuscript is as follows. In Section \ref{sec:kernels}, we recall the evolution equations for the different parton distributions and the corresponding kernels, introducing the notation required to present our results. In Section \ref{sec:results}, we summarize the method used to obtain the correction to the splitting functions and present the corresponding kernels. Finally, in Section \ref{sec:conclusions}, we expose our conclusions.

\section{Splitting kernels and parton distribution basis}
\label{sec:kernels}
We start by writing down the general expression for the evolution of gluon, photon and quark distributions as \cite{Ellis:1996nn}
\beqn
\frac{dg}{dt}&=& \sum_{j=1}^{n_F} P_{g q_j} \otimes  q_j+ \sum_{j=1}^{n_F} P_{g \bar{q}_j} \otimes \bar{q}_j +  P_{g g} \otimes g +  P_{g \gamma} \otimes \gamma \, ,
\\ \frac{d\gamma}{dt}&=& \sum_{j=1}^{n_F} P_{\gamma q_j} \otimes  q_j+ \sum_{j=1}^{n_F} P_{\gamma \bar{q}_j} \otimes \bar{q}_j + P_{\gamma g} \otimes g + P_{\gamma \gamma} \otimes \gamma \, ,
\eeqn
\beqn
\frac{dq_i}{dt}&=& \sum_{j=1}^{n_F} P_{q_i q_j} \otimes q_j + \sum_{j=1}^{n_F} P_{q_i \bar{q}_j} \otimes \bar{q}_j + P_{q_i g} \otimes g + P_{q_i \gamma} \otimes \gamma \, ,
\eeqn
with $t=\ln{\mu^2}$ ($\mu$ being the factorization scale) and $P_{ij}$ the Altarelli-Parisi splitting functions in the space-like region. Evolution equations for antiquarks can be obtained by applying conjugation invariance. Here we use the notation
\beq
(f\otimes g)(x) = \int_x^1 \, \frac{dy}{y} \, f\left(\frac{x}{y}\right)g(y) \, ,
\eeq
to indicate convolutions. We do not include the lepton distributions in this work, since up to the order we reach here they basically factorize from the rest of the distributions \footnote{To $\cal{O}(\alpha)$ lepton distributions only couple, in a trivial way, to the photon density.}.
Along this work we will present the expressions for the splitting functions including QCD and QED corrections. Each kernel can be expanded as
\beqn
P_{ij} &=& \as P_{ij}^{(1,0)} +\as^2 P_{ij}^{(2,0)}+ \as^3 P_{ij}^{(3,0)} + a P_{ij}^{(0,1)} + \as \, a P_{ij}^{(1,1)} + ... \, ,
\eeqn
where the upper indices indicate the (QCD,QED) order of the calculation, with $\as \equiv \frac{\alphas}{2\pi}$ and $a \equiv \frac{\alpha}{2\pi}$. Due to the QED corrections, the Altarelli-Parisi splitting kernels can depend on the electric charge of the initiating quarks (up or down type), such that in general $P_{q}^{(n,1)} \sim e_q^2$.

The quark splitting functions are decomposed as
\beqn
P_{q_i \, q_k} &=& \delta_{ik} \, P^V_{qq} + P^S_{qq} \, ,\\
 P_{q_i \, \bar{q}_k} &=& \delta_{ik} \, P^V_{q\bar{q}} + P^S_{q\bar{q}} \, ,\\
  P_q^{\pm} &=& P^V_{qq} \pm P^V_{q\bar{q}} \, ,
\eeqn
which acts as a definition for $P^V_{q q}$ and $P^V_{q \bar{q}}$. In order to minimize the mixing between the different parton distributions in the evolution, it is convenient to introduce the following basis \cite{Roth:2004ti}:
\beq
\label{eq:basis}
\{u_v,d_v,s_v,c_v,b_v,
\Delta_{uc},\Delta_{ds},\Delta_{sb},\Delta_{UD}, \Sigma,g,\gamma\} \, ,
\eeq
where
\beqn
q_{v_i} &=& q_i-\bar{q_i} \, , \\
\Delta_{uc} &=& u+\bar{u}-c-\bar{c} \, ,\nn \\ 
\Delta_{ds} &=& d+\bar{d}-s-\bar{s} \, ,\nn \\ 
\Delta_{sb} &=& s+\bar{s}-b-\bar{b} \, , \\ 
\Delta_{UD} &=& u+\bar{u}+c+\bar{c} -d-\bar{d} -s-\bar{s}-b-\bar{b} \, , \\
 \Sigma &=& \sum_{i=1}^{n_F} ( q_i+\bar{q}_i)  \, .
\eeqn
$\Delta_{UD} $ could also include the top quark distribution in case of a 6 flavour analysis (adding $\Delta_{ct}$ and $t_v$ to complete the basis). Identical results are obtained by using a similar basis proposed in Ref. \cite{Bertone:2015lqa}. In the evolution equations for the corresponding distribution, we do take into account that beyond NLO in QCD the singlet  {\it non-diagonal terms} ($P^S_{q \bar{q}}$ and $P^S_{qq}$) are different \cite{Catani:2004nc}. Hence, it is useful to define 
\beqn
\Delta P^S & \equiv & P^S_{qq} -P^S_{q \bar{q}}  , \nn \\
 P^S & \equiv &  P^S_{qq} + P^S_{q \bar{q}} , 
\eeqn
where we explicitly use that these contributions do not depend on the quark charge up to the order we reach here, since they do not receive QED corrections to $\cal{O}(\alpha)$.

The evolution equations for the parton distributions in the basis of Eq.(\ref{eq:basis}) read,
\beqn
\frac{dq_{v_i}}{dt} &=&   P_{q_i}^-     \otimes q_{v_i}  +\sum_{j=1}^{n_F} \Delta P^S  \otimes  q_{v_j}    \, ,
\label{eq:evolucionqvSIMPLE}
\\ \frac{d \{ \Delta_{uc} , \Delta_{ct} \}}{dt} &=&  P_{u}^+ \otimes \{ \Delta_{uc} , \Delta_{ct}  \} \, ,
\label{eq:evolucionDupperSIMPLE}
\\ \frac{d \{ \Delta_{ds} ,\Delta_{sb}  \}}{dt} &=&  P_{d}^+ \otimes \{ \Delta_{ds} ,\Delta_{sb}  \} \, ,
\label{eq:evolucionDlowerSIMPLE}
\\ \nn \frac{d  \Delta_{UD}  }{dt} &=&  \frac{P_{u}^+ + P_{d}^+}{2} \otimes \Delta_{UD} 
                                                  + \frac{P_{u}^+ - P_{d}^+}{2} \otimes \Sigma +  (n_u-n_d) P^S \, \otimes \Sigma
                                                  \\ &+& 2 (n_u P_{ug} -n_d P_{dg})  \otimes g + 2 (n_u P_{u\gamma} -n_d P_{d\gamma})  \otimes \gamma \, ,
                                                  \label{eq:evolucionDUDSIMPLE}
\\ \nn 
\frac{d  \Sigma  }{dt} &=&  \frac{P_{u}^+ + P_{d}^+}{2} \otimes \Sigma
                                                  + \frac{P_{u}^+ - P_{d}^+}{2} \otimes \Delta_{UD} +  n_F \, P^S \otimes \Sigma
                                                  \\ &+& 2 (n_u P_{ug} +n_d P_{dg})  \otimes g + 2 (n_u P_{u\gamma} +n_d P_{d\gamma})  \otimes \gamma \, .
                                                  \label{eq:evolucionSIGMASIMPLE}
\eeqn
Notice that in the limit of equal number of $u$ and $d$ quarks ($n_u=n_d$) and same electric charges ($P_{ug}=P_{dg}, P_{u\gamma}=P_{d\gamma}$), $\Delta_{UC}$ decouples from the other distributions in the evolution, while the singlet evolution recovers the usual pure-QCD expression.

\section{QCD-QED splitting kernels}
\label{sec:results}
To set the correct normalization, we start by reminding the lowest order splitting functions in QCD $P_{ij}^{(1,0)}$ \cite{Altarelli:1977zs}
\beqn
\nn P_{qq}^{(1,0)}(x) &=& C_F \left[  \frac{1+x^2}{(1-x)_+}  +\frac{3}{2} \delta (1-x) \right] = C_F \, p_{qq}(x) + \frac{3 \, C_F}{2} \delta(1-x) \, , 
\\ \nn P_{qg}^{(1,0)}(x) &=& T_R \left[   x^2+(1-x)^2\right]= T_R \, p_{qg}(x) \, ,
\\ \nn P_{gq}^{(1,0)}(x) &=& C_F \left[   \frac{1+(1-x)^2}{x}  \right]= C_F \, p_{gq}(x) \, , 
\eeqn
\beqn
P_{gg}^{(1,0)}(x) &=& 2 C_A \left[  \frac{x}{(1-x)_+}   +\frac{1-x}{x} + x(1-x) \right] +  \frac{\beta_0}{2} \delta(1-x) \, ,
\eeqn
with $\beta_0=\frac{11 N_C-4 n_F T_R}{3}$ and the usual plus distribution defined as
\beq
\int_0^1 dx \, \frac{f(x)}{(1-x)_+} = \int_0^1 dx \, \frac{f(x)-f(1)}{1-x} \, ,
\eeq
for any regular test function $f$.
In the same way, the lowest order splitting functions in QED $P_{ij}^{(0,1)}$ are given by \cite{Roth:2004ti}
\beqn
P_{qq}^{(0,1)}(x) &=& e_q^2 \left[  p_{qq}(x) +\frac{3}{2} \delta (1-x) \right] \, , \nn \\
P_{q\gamma}^{(0,1)}(x) &=& N_C \, e_q^2  \, p_{qg}(x) \, ,  \nn \\
P_{\gamma q}^{(0,1)}(x) &=&  e_q^2 \, p_{gq}(x) \, , \nn \\
P_{\gamma\gamma}^{(0,1)}(x) &=& -\frac{2}{3} \sum_{f} e_f^2 \, \delta (1-x)  \, ,
\eeqn
where there is an explicit dependence on the quark electromagnetic (EM) charge. Furthermore, the sum over fermion charges in the $P^{(0,1)}_{\gamma\gamma}$ kernel corresponds to the definition 
\beq
\sum_{f}e_f^2 =N_C \sum_{q}^{n_F}e_q^2 \, + \, \sum_{l}^{n_L} e_l^2 \, ,
\eeq
with $n_F$ and $n_L$ the number of quark and lepton flavours, respectively.

The expressions for NLO QCD corrections to the splitting functions $P_{ij}^{(2,0)}$ can be found in Refs. \cite{Curci:1980uw,Furmanski:1980cm,Ellis:1996nn} while the NNLO ones $P_{ij}^{(3,0)}$ are in Ref. \cite{Moch:2001im,Moch:2004pa,Vogt:2004mw,Vogt:2005dw}. Moreover, NLO QCD corrections to the splitting functions with photons are also available, both at amplitude and squared-amplitude level \cite{Gluck:1983mm,Gluck:1991ee,Fontannaz:1992gj,Sborlini:2013jba}.
In order to obtain the mixed NLO QCD-QED corrections $P_{ij}^{(1,1)}$, we start by analyzing the computation of the two loop anomalous dimensions in the light-cone gauge, originally performed for the non-singlet component by Curci, Furmanski and Petronzio in Ref. \cite{Curci:1980uw} and extended to the singlet case in Ref. \cite{Furmanski:1980cm,Ellis:1996nn}. Roughly speaking, $P_{ij}^{(1,1)}$ can be obtained from $P_{ij}^{(2,0)}$ by carefully taking a particular Abelian limit, i.e. by replacing one gluon by a photon \cite{Kataev:1992dg}. While in principle the limit is straightforward, there are some particularities that can lead to misleading results. To avoid that, we strongly rely on the detailed documentation presented for the non-singlet in Ref. \cite{Curci:1980uw} and for the singlet in Ref. \cite{Ellis:1996nn}, where results for each topological contribution and the corresponding colour factor are carefully registered. Therefore, we recompute the colour factor for each contribution by selecting only those that are relevant for the NLO QCD-QED mixed terms. 

The introduction of a photon is not only associated with the corresponding Abelian limit, but also involves the need of a charge separation. We consider quarks of flavour $q$ with electric charge $e_u=2/3$ or $e_d=-1/3$. The contribution of each quark flavour is individualized by \emph{carefully} considering the limit $n_F \to 1$. Nevertheless, the potential presence of internal quark loops forces us to distinguish between real and virtual $n_F$ contributions. 

\begin{figure}[htb]
\begin{center}
\includegraphics[width=0.53\textwidth]{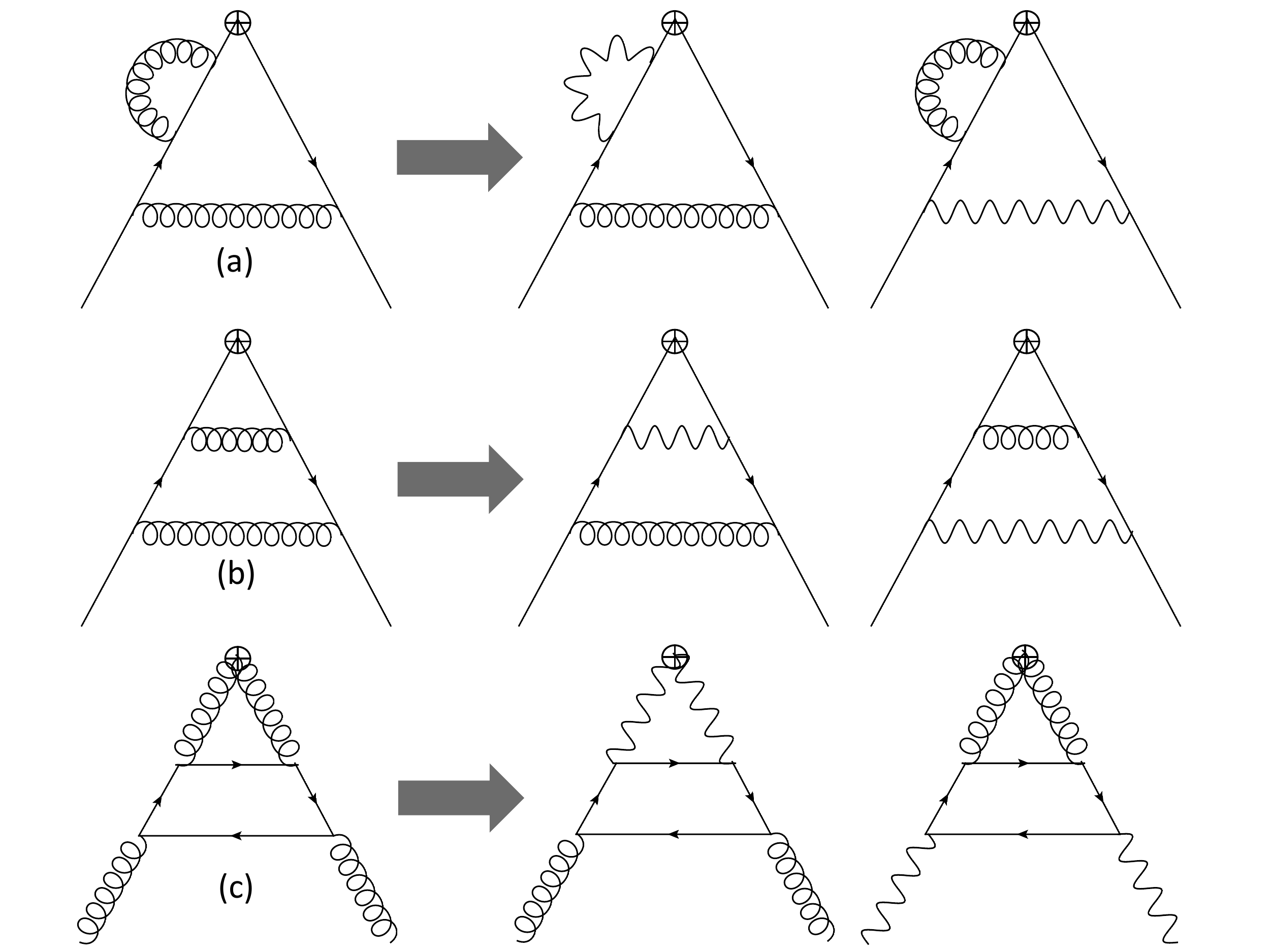}
\caption{A sample of diagrams associated with the virtual and real contributions to $P^{(2,0)}_{qq}$, in (a) and (b) respectively. To obtain $P^{(1,1)}_{qq}$, one gluon is replaced by a photon. Since there are two ways to perform the replacement, a factor $2$ arises. In (c), $P_{gg}^{(2,0)}$ is considered with a representative diagram. In this case, the Abelian limit allows to compute both $P^{(1,1)}_{\gamma g}$ and $P^{(1,1)}_{g \gamma}$. The presence of a fermionic box forces to take into account the different quark EM charges.}
\label{fig:colourTratamiento}
\end{center}
\end{figure}

Let's describe the algorithm that allows to obtain the QED corrections by {\rm replacing gluons by photons} from the QCD splitting functions.
\begin{enumerate}
	\item Since the QCD kernels include the average over initial colour states, we first correct the overall normalization of $P_{ba}$ in the case that an initial gluon ($a=g$) has to be transformed into a photon ($a=\gamma$),  multiplying the kernel by $(N_C^2-1$).
  \item Then, we identify those Feynman diagrams that are non-vanishing when replacing the corresponding gluon by a photon, and compute their colour structure. If the original QCD diagram involves two non-observable gluons, the replacement $g \to \gamma$ leads to two non-equivalent topologies (both in real and virtual terms). At ${\cal O}(\alpha \, \alphas)$, it is necessary to multiply the final result by a global factor $2$ to account for this effect in the pure quark kernels.
  \item After that, we write the colour structures in terms of $N_C$ by using the well known relations
\beq
C_A=N_C \ \ \ \ , \ \ \ \ C_F=-\frac{1}{2 \, N_C} + \frac{N_C}{2} \ .
\eeq
  \item Next, we single out and keep only the leading terms in the limit $N_C \to 0$.
  \item The final step consists in recomputing the colour structure for the Abelian diagrams, replacing the QCD ones  in the expression of $P_{ba}$. 
  \end{enumerate}
In practical terms, we notice that at this order the QED results can be recovered by simply identifying the most divergent colour structure and performing the replacement directly there, with the additional normalization change if the initial gluon is replaced by a photon or if there are two unresolved gluons. Finally, if the Feynman diagram expansion involves fermion loops, we use the replacement
\beqn
n_F &\to& \sum_{j=1}^{n_F} e_{q_j}^2 \, ,
\label{eq:ReemplazoNF}
\eeqn
whilst for external quarks we just multiply the result by the global factor $e_q^2$. 
Fig. \ref{fig:colourTratamiento} provides a graphical representation of the Abelianization algorithm applied to the NLO QCD splitting kernels to obtain the mixed QCD-QED corrections. In particular, in (c), we explicitly motivate the replacement rule mentioned in \Eq{eq:ReemplazoNF} by exploring a typical contribution to $P^{(2,0)}_{gg}$. When one gluon is replaced by a photon, we obtain a fermion box with two photons attached to it; the QED interaction introduces a factor $e_q^2$ responsible of a charge separation for each quark flavour.

In the context of the full EW theory, the corrections induced by massive bosons lead to singularities. However, we will not deal with them in this work because it is possible to factorize them and achieve a fully consistent treatment of IR divergences relying only on QCD-QED splittings. In other terms, singularities introduced by $W$ and $Z$ bosons can be absorbed into the hard scattering, thus leaving unaffected the evolution of PDFs.

We therefore present the $({\rm QCD,QED})=(1,1)$ expressions of the corresponding splitting kernels. In first place, we obtain
\beqn
\nn P_{q\gamma}^{(1,1)} &=& \frac{C_F \, C_A \, e_q^2}{2} \left\{ 4 -9x-(1-4x) \ln{x} - (1-2x) \vphantom{\lnn{\frac{1-x}{x}}} \lnn{x} +4 \ln{1-x} \right.
\\ &+& \left. \vphantom{\lnn{\frac{1-x}{x}}} p_{qg}(x) \left[  2\lnn{\frac{1-x}{x}} - 4\ln{\frac{1-x}{x}}  -\frac{2\pi^2}{3}+10 \right]
 \right\}  \, ,
 \\ P_{g\gamma}^{(1,1)} &=& C_F \, C_A \, \left(\sum_{j=1}^{n_F} e_{q_j}^2\right) \, \left\{ -16 +8 x +\frac{20}{3}x^2+\frac{4}{3x} - \vphantom{\frac{20}{3}x^2} (6+10x) \ln{x} -2(1+x)\lnn{x}  \right\}  , \
 \\ P_{\gamma\gamma}^{(1,1)} &=& -C_F \, C_A \, \left(\sum_{j=1}^{n_F} e_{q_j}^2\right)  \delta(1-x) \, ,
  \eeqn
for photon initiated processes, and
\beqn
\nn P_{qg}^{(1,1)} &=&  \frac{T_R \, e_q^2}{2} \left\{ 4 -9x-(1-4x) \ln{x} - (1-2x) \vphantom{\lnn{\frac{1-x}{x}}} \lnn{x} +4 \ln{1-x} \right.
\\ &+& \left. \vphantom{\lnn{\frac{1-x}{x}}} p_{qg}(x) \left[  2\lnn{\frac{1-x}{x}} - 4\ln{\frac{1-x}{x}}  -\frac{2\pi^2}{3}+10 \right]
 \right\}\, , 
 \\ P_{\gamma g}^{(1,1)} &=& T_R \,\left(\sum_{j=1}^{n_F} e_{q_j}^2\right) \,  \left\{ -16 +8 x +\frac{20}{3}x^2+\frac{4}{3x} - \vphantom{\frac{20}{3}x^2} (6+10x) \ln{x} -2(1+x)\lnn{x}  \right\} \, , 
 \\ P_{gg}^{(1,1)} &=& -T_R \,\left(\sum_{j=1}^{n_F} e_{q_j}^2\right) \, \delta(1-x) \, ,
 \eeqn
for collinear splitting processes with a starting gluon. Notice that QED corrections to the diagonal splitting kernels $P_{\gamma \gamma}^{(1,1)}$ and $P_{g g}^{(1,1)}$ are proportional to the Dirac's delta function $\delta(1-x)$ since they are originated by virtual two-loop  contributions to the photon and gluon propagators, respectively. On the other hand, the quark splitting functions are given by
\beqn
P_{qq}^{S(1,1)} &=& P_{q\bar{q}}^{S(1,1)}=0 \, , 
\\ \nn P_{qq}^{V(1,1)} &=& -  2 \, C_F \, e_q^2 \left[\left(2 \ln{1-x}+\frac{3}{2}\right)\ln{x} p_{qq}(x) + \frac{3+7x}{2}\ln{x} + \frac{1+x}{2}{\lnn{x}}  \right.
\\ &+& \left. 5(1-x) + \left( \frac{\pi^2}{2}-\frac{3}{8}-6 \zeta_3 \right) \delta(1-x)  \right] \, , 
\\ P_{q\bar{q}}^{V(1,1)} &=&  2\, C_F \, e_q^2 \left[4(1-x)+2(1+x)\ln{x} + 2p_{qq}(-x)S_2(x)\right] \, , 
\\ \nn P_{gq}^{(1,1)} &=& C_F \, e_q^2 \left[-(3\ln{1-x}+\lnn{1-x})p_{gq}(x) \vphantom{\left(2+\frac{7}{2}x\right)} + \left(2+\frac{7}{2}x\right)\ln{x}\right.
\\ &-& \left. \left(1-\frac{x}{2}\right)\lnn{x}  - 2x\ln{1-x}-\frac{7}{2}x-\frac{5}{2}\right] \, , 
\\ P_{\gamma q}^{(1,1)} &=& P_{g q}^{(1,1)} \, ,
\eeqn
where we appreciate that singlet contributions vanish at this order, as anticipated in Sec. \ref{sec:kernels}. The function $S_2(x)$ is given by
\beqn
\nn S_2(x) &=& \int_{\frac{x}{1+x}}^{\frac{1}{1+x}} \, \frac{dz}{z} \, \ln{\frac{1-z}{z}} = \li{-\frac{1}{x}} - \li{-x}
\\ &+&  \lnn{\frac{x}{1+x}}-\lnn{\frac{1}{1+x}}\,.
\label{eq:S2definicion}
\eeqn

Finally, we establish the consistency of our results by checking the corresponding fermionic and momentum sum rules for each distribution. Explicitly, the ${\cal O}(\alpha \, \alphas)$ contributions to the evolution kernels fulfill:
\beqn
\int_0^1 \, dx \, && \left(P_{qq}^{V(1,1)}-P_{q\bar{q}}^{V(1,1)} \right) = 0 \, ,
\\ \int_0^1 \, dx  &x& \left(2\, n_u P_{u g}^{(1,1)} + 2\,  n_d P_{d g}^{(1,1)} + P_{\gamma g}^{(1,1)} + P_{g g}^{(1,1)}  \right) = 0 \, ,
\\ \int_0^1 \, dx &x& \left(2\,  n_u P_{u \gamma}^{(1,1)} +2\, n_d P_{d \gamma}^{(1,1)} + P_{g \gamma}^{(1,1)} + P_{\gamma \gamma}^{(1,1)}  \right) = 0 \, ,
\\ \int_0^1 \, dx  &x& \left(P_{q q}^{V(1,1)} + P_{q \bar{q}}^{V(1,1)} + P_{g q}^{(1,1)} + P_{\gamma q}^{(1,1)}  \right)  = 0 \, .
\eeqn

\section{Conclusions}
\label{sec:conclusions}
In this article, we discussed the computation of the NLO mixed QCD-QED corrections to the Altarelli-Parisi evolution kernels. 
In order to reach that accuracy, we analyzed the colour structure of each diagram contributing to these corrections and evaluated their modification after a gluon is transformed into a photon. Then, we computed the explicit expressions for the evolution kernels by carefully considering the Abelian limit of the results available in the literature for pure QCD processes. In particular, relying on Refs. \cite{Curci:1980uw,Furmanski:1980cm,Ellis:1996nn} we obtained the corresponding results up to ${\cal O}(\alpha \, \alphas)$.

The computation of higher-order mixed QCD-QED contributions to physical observables plays a crucial role in the full program of precision computations for hadron colliders. In this direction, the results provided here are useful to improve the accuracy of the PDFs sets used to perform the theoretical predictions required by nowadays (and future) experiments.

\begin{acknowledgements}
This work is partially supported by UBACYT, CONICET, ANPCyT,
by the Spanish Government and EU ERDF funds
 (grants FPA2014-53631-C2-1-P, FPA2011-23778 and SEV-2014-0398) and by GV (PROMETEUII/2013/007).
\end{acknowledgements}



\end{document}